\documentclass{PoS}

\def\dd{\displaystyle}

\def\bea{\begin{eqnarray}}
\def\eea{\end{eqnarray}}
\def\beq{\begin{equation}}
\def\eeq{\end{equation}}
\def\bq{\begin{quote}}
\def\eq{\end{quote}}
\def\be{\begin{equation}}
\def\ee{\end{equation}}
\def\bc{\begin{center}}
\def\ec{\end{center}}
\def\bea{\begin{eqnarray}}
\def\eea{\end{eqnarray}}
\def\dd{\displaystyle}

\def\GeV{{\rm GeV}}
\def\gappeq{\mathrel{\rlap {\raise.5ex\hbox{$>$}} {\lower.5ex\hbox{$\sim$}}}}
\def\lappeq{\mathrel{\rlap{\raise.5ex\hbox{$<$}} {\lower.5ex\hbox{$\sim$}}}}

\title{Status of Neutrino Masses and Mixing in 2010}

\ShortTitle{Neutrino Masses}

\author{\speaker{Guido Altarelli}\thanks{This work has been partly supported by the Italian Ministero 
dell'Universit\`a e della Ricerca Scientifica, under the COFIN program (PRIN 2008) 
and by the European Commission
under the network  "Heptools"}\\
        Dipartimento di Fisica `E.~Amaldi', Universit\`a di Roma Tre
and INFN, Sezione di Roma Tre, I-00146 Rome, Italy and \\ CERN, Department of Physics, Theory Unit, 
 CH--1211 Geneva 23, Switzerland\\
        E-mail: \email{guido.altarelli@cern.ch}}

\abstract{We present a short summary of our present knowledge and understanding of neutrino masses and mixing.
}

\FullConference{Quarks, Strings and the Cosmos - Hector Rubinstein Memorial Symposium\\
		August 09-11, 2010\\
		AlbaNova, Stockholm, Sweden
		\\
RM3-TH/10-12,                 CERN-PH-TH/2010-280
\\
}

\begin{document}

\section{Experimental facts}
Experiments on neutrino oscillations, which measure differences of squared masses and mixing angles \cite{review}, \cite{rev2} have established that neutrinos have a mass. Two distinct oscillation frequencies have been first measured in solar and atmospheric neutrino oscillations and later confirmed by experiments on earth, like KamLAND, K2K and MINOS. A signal corresponding to a third mass difference was claimed by the LSND experiment (with antineutrinos) but not confirmed by KARMEN. More recently MiniBooNE \cite{Mini} has reported some possible supporting evidence for the LSND effect in their antineutrino run while no oscillation is observed in the neutrino run. Two well separated differences need at least three different neutrino mass eigenstates involved in oscillations so that the three known neutrino species can be sufficient. Then at least two $\nu$'s must be massive while, in principle, the third one could still be massless. The existence of a third oscillation frequency would imply the need for additional sterile neutrinos (i.e. with no weak interactions, as any new light active neutrino was excluded by LEP) or CPT violation (as, in this case, the masses of neutrinos and antineutrinos can be different). The MiniBooNE experiment is continuing to take data and it is very interesting to see whether the hint for a new frequency will be confirmed. In the following we will assume the simplest picture with no new frequency, three active neutrinos, no sterile neutrinos and CPT invariance. The mass eigenstates involved in solar oscillations are $m_1$ and $m_2$ and, by definition, $|m_2|> |m_1|$, so that $\Delta m^2_{sun}=|m_2|^2-|m_1|^2>0$. The atmospheric neutrino oscillations involve $m_3$:  $\Delta m^2_{atm}=|\Delta m^2_{31}|$ with $\Delta m^2_{31}=|m_3|^2-|m_1|^2$ either positive (normal hierarchy) or negative (inverse hierarchy). The present data are still compatible with both cases. The degenerate spectrum occurs when the average absolute value of the masses is much larger than all mass squared differences: $|m_i|^2 >> |\Delta m^2_{hk}|$. With the standard set of notations and definitions \cite{review} the present data are summarised in Table(\ref{tab:data}).

\begin{table}[h]
\begin{center}
\begin{tabular}{|c|c|c|}
  \hline
  Quantity & ref. \cite{FogliIndication} & ref. \cite{MaltoniIndication} \\
  \hline
  $\Delta m^2_{sun}~(10^{-5}~{\rm eV}^2)$ &$7.67^{+0.16}_{-0.19}$ & $7.59^{+0.23}_{-0.20}$  \\
  $\Delta m^2_{atm}~(10^{-3}~{\rm eV}^2)$ &$2.39^{+0.11}_{-0.08}$ & $2.40^{+0.12}_{-0.11}$  \\
  $\sin^2\theta_{12}$ &$0.312^{+0.019}_{-0.018}$ & $0.318^{+0.019}_{-0.016}$ \\
  $\sin^2\theta_{23}$ &$0.466^{+0.073}_{-0.058}$ &  $0.50^{+0.07}_{-0.06}$ \\
  $\sin^2\theta_{13}$ &$0.016\pm0.010$ &$0.013^{+0.013}_{-0.009}$  \\
  \hline
  \end{tabular}
\end{center}
\begin{center}
\begin{minipage}[t]{12cm}
\caption{\label{tab:data} Fits to neutrino oscillation data.} 
\end{minipage}
\end{center}
\end{table}

Oscillation experiments do not provide information about the absolute neutrino mass scale. Limits on that are obtained \cite{review} from the endpoint of the tritium beta decay spectrum, from cosmology and from neutrinoless double beta decay ($0\nu \beta \beta$). From tritium we have an absolute upper limit of
2.2 eV (at 95\% C.L.) on the mass of electron  antineutrino, which, combined with the observed oscillation
frequencies under the assumption of three CPT-invariant light neutrinos, also amounts to an upper bound on the masses of
the other active neutrinos. Complementary information on the sum of neutrino masses is also provided by the galaxy power
spectrum combined with measurements of the cosmic  microwave background anisotropies. According to recent analyses of the most reliable data \cite{fo}
$\sum_i \vert m_i\vert < 0.60\div 0.75$ eV (at 95\% C.L.) depending on the retained data (the number for the sum has to be divided by 3 in order to obtain a limit on the mass of each neutrino).
The discovery of $0\nu \beta \beta$ decay would be very important because it would establish lepton number violation and
the Majorana nature of $\nu$'s, and provide direct information on the absolute
scale of neutrino masses.
The present limit from $0\nu \beta \beta$  (with large ambiguities from nuclear matrix elements) is about $\vert m_{ee}\vert < (0.3\div 0.8)$ eV \cite{fo} (see eq. (\ref{3nu1gen})). 
 
\section{Majorana Neutrinos and the See-Saw Mechanism}

Given that neutrino masses are certainly extremely
small, it is really difficult from the theory point of view to avoid the conclusion that the lepton number L conservation is probably violated and that $\nu$'s are Majorana fermions.
In this case the smallness of neutrino masses can naturally be explained as inversely proportional
to the very large scale where L conservation is violated, of the order of the grand unification scale $M_{GUT}$ or maybe, for the lightest among them, the Planck scale $M_{Pl}$ \cite{seesaw}. 
If neutrinos are Majorana particles, their masses arise from the generic dimension-five non renormalizable operator of the form: 
\be
O_5=\frac{(H l)^T_i \lambda_{ij} (H l)_j}{M}+~h.c.~~~,
\label{O5}
\ee  
with $H$ being the ordinary Higgs doublet, $l_i$ the SU(2) lepton doublets, $\lambda$ a matrix in  flavour space,
$M$ a large scale of mass and a charge conjugation matrix $C$
between the lepton fields is understood. 

Neutrino masses generated by $O_5$ are of the order
$m_{\nu}\approx v^2/M$ for $\lambda_{ij}\approx {\rm O}(1)$, where $v\sim {\rm O}(100~\GeV)$ is the vacuum
expectation value of the ordinary Higgs. In the simplest case the exchanged particle is the right-handed (RH) neutrino $\nu^c$ (a gauge singlet fermion here described through its charge conjugate field), and the resulting neutrino mass matrix reads (type I see-saw \cite{seesaw}):
\be  
m_{\nu}=m_D^T M^{-1}m_D~~~.
\ee 
where $m_D$ and $M$ denote the Dirac neutrino mass matrix (defined as $ {\nu^c}^T m_D \nu$) and the Majorana mass matrix of $\nu^c$ (defined as $ {\nu^c}^T M \nu^c$), respectively.
As one sees,  the light neutrino masses are quadratic in the Dirac
masses and inversely proportional to the large Majorana mass.  For
$m_{\nu}\approx \sqrt{\Delta m^2_{atm}}\approx 0.05$ eV and 
$m_{\nu}\approx m_D^2/M$ with $m_D\approx v
\approx 200$~GeV we find $M\approx 10^{15}$~GeV which indeed is an impressive indication that the scale for lepton number violation is close to
$M_{GUT}$. Thus probably neutrino masses are a probe into the physics near $M_{GUT}$. This argument, in my opinion, strongly discourages models where neutrino masses are generated near the weak scale and are suppressed by some special mechanism.

\section{Importance of Neutrinoless Double Beta Decay}

Oscillation experiments cannot distinguish between
Dirac and Majorana neutrinos.
The detection of neutrino-less double beta decay would provide direct evidence of $L$ non conservation, and the Majorana nature of neutrinos. It would also offer a way to possibly disentangle the 3 cases of degenerate, normal or inverse hierachy neutrino spectrum.  The quantity which is bound by experiments on $0\nu \beta \beta$
is the 11 entry of the
$\nu$ mass matrix, which in general, from $m_{\nu}=U^* m_{diag} U^\dagger$, is given by :
\bea 
\vert m_{ee}\vert~=\vert(1-s^2_{13})~(m_1 c^2_{12}~+~m_2 s^2_{12})+m_3 e^{2 i\phi} s^2_{13}\vert
\label{3nu1gen}
\eea
where $m_{1,2}$ are complex masses (including Majorana phases) while $m_3$ can be taken as real and positive and $\phi$ is the $U_{PMNS}$ phase measurable from CP violation in oscillation experiments. Starting from this general formula it is simple to
derive the bounds for degenerate, inverse hierarchy or normal hierarchy mass patterns shown in Fig.1 \cite{fsv}.

\begin{figure}
\centering
\includegraphics[width=10cm]{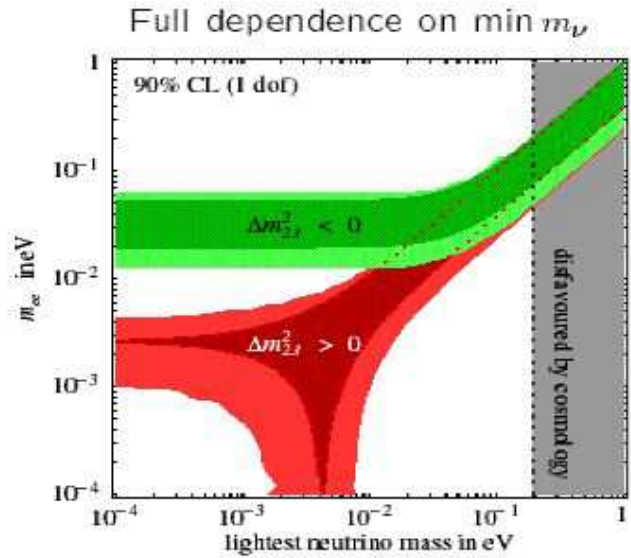}    
\caption[ ]{A plot \cite{fsv} of $m_{ee}$ in eV, the quantity measured in neutrino-less double beta decay, given in eq.(\ref{3nu1gen}), versus the lightest neutrino mass $m_1$, also in eV. The upper (lower) band is for inverse (normal) hierarchy.}
\end{figure}

In the next few years a new generation of experiments will reach a larger sensitivity on $0\nu \beta \beta$ by about an order of magnitude. If these experiments will observe a signal this would indicate that the inverse hierarchy is realized, if not, then the normal hierarchy case remains a possibility. 

\section{Baryogenesis via Leptogenesis from Heavy $\nu^c$ Decay}

In the Universe we observe an apparent excess of baryons over antibaryons. It is appealing that one can explain the
observed baryon asymmetry by dynamical evolution (baryogenesis) starting from an initial state of the Universe with zero
baryon number.  For baryogenesis one needs the three famous Sakharov conditions: B violation, CP violation and no thermal
equilibrium. In the history of the Universe these necessary requirements have possibly occurred at different epochs. Note
however that the asymmetry generated during one epoch could be erased in following epochs if not protected by some dynamical
reason. In principle these conditions could be fulfilled in the SM at the electroweak phase transition. In fact, when kT is of the order of the weak scale, B conservation is violated by
instantons (but B-L is conserved), CP symmetry is violated by the CKM phase and
sufficiently marked out-of- equilibrium conditions could be realized during the electroweak phase transition. So the
conditions for baryogenesis  at the weak scale in the SM superficially appear to be present. However, a more quantitative
analysis
\cite{tro} shows that baryogenesis is not possible in the SM because there is not enough CP violation and the phase
transition is not sufficiently strong first order, unless the Higgs mass is below a bound which by now is excluded by LEP. In SUSY extensions of the SM, in particular in the MSSM,
there are additional sources of CP violation and the bound on $m_H$ is modified but also this possibility has by now become at best marginal after the results from LEP2.

If baryogenesis at the weak scale is excluded by the data it can occur at or just below the GUT scale, after inflation.
But only that part with
$|{\rm B}-{\rm L}        |>0$ would survive and not be erased at the weak scale by instanton effects. Thus baryogenesis at
$kT\sim 10^{10}-10^{15}~{\rm GeV}$ needs B-L violation and this is also needed to allow $m_\nu$ if neutrinos are Majorana particles.
The two effects could be related if baryogenesis arises from leptogenesis then converted into baryogenesis by instantons
\cite{buch}. The decays of heavy Majorana neutrinos (the heavy eigenstates of the see-saw mechanism) happen with violation of lepton number L, hence also of B-L and can well involve a sufficient amount of ¤CP violation. Recent results on neutrino masses are compatible with this elegant possibility. Thus the case
of baryogenesis through leptogenesis has been boosted by the recent results on neutrinos.

\section{Models of Neutrino Mixing}

By now, after KamLAND, SNO and the upper limits on the absolute value of neutrino masses, not too much hierarchy in the spectrum of neutrinos is indicated by experiments: 
\bea
r = \Delta m_{sol}^2/\Delta m_{atm}^2 \sim 1/30.\label{r}
\eea
Precisely $r=0.032^{+0.006}_{-0.005}$ at $3\sigma$'s  \cite{FogliIndication,MaltoniIndication}. Thus, for a hierarchical spectrum, $m_2/m_3 \sim \sqrt{r} \sim 0.2$, which is comparable to the Cabibbo angle $\lambda_C \sim 0.22$ or $\sqrt{m_{\mu}/m_{\tau}} \sim 0.24$. This suggests that the same hierarchy parameter (raised to powers with o(1) exponents) may apply for quark, charged lepton and neutrino mass matrices. This in turn indicates that, in the absence of some special dynamical reason, we do not expect quantities like $\theta_{13}$ or the deviation of  $\theta_{23}$ from its maximal value to be too small. Indeed it would be very important to know how small the mixing angle $\theta_{13}$  is and how close to maximal $\theta_{23}$ is. 

Neutrino mixing is important because it could in principle provide new clues for the understanding of the flavour problem. Even more so since neutrino mixing angles show a pattern that is completely different than that of quark mixing: for quarks all mixing angles are small, for neutrinos two angles are large (one is even compatible with the maximal value) and only the third one is small. For building up theoretical models of neutrino mixing one must guess which features of the data are really relevant in order to identify the basic principles for the formulation of the model.  We see from Table(\ref{tab:data})  \cite{FogliIndication,MaltoniIndication} that within measurement errors
the observed neutrino mixing matrix is compatible with
the so called Tri-Bimaximal (TB) form \cite{hps}. The best measured neutrino mixing angle $\theta_{12}$ is just about 1$\sigma$ below the TB value $\sin^2{\theta_{12}}=1/3$, while the maximal value for $\theta_{23}$ is well inside the 1-$\sigma$ interval and $\theta_{13}$ is still compatible with zero (see Table \ref{tab:data}). In fact, the  TB mixing matrix (in a particular phase convention) is given by:
\begin{equation}
U_{TB}= \left(\matrix{
\dd\sqrt{\frac{2}{3}}&\dd\frac{1}{\sqrt 3}&0\cr
-\dd\frac{1}{\sqrt 6}&\dd\frac{1}{\sqrt 3}&-\dd\frac{1}{\sqrt 2}\cr
-\dd\frac{1}{\sqrt 6}&\dd\frac{1}{\sqrt 3}&\dd\frac{1}{\sqrt 2}}\right)~~~~~. 
\label{2}
\end{equation}

Thus, one possibility is that one takes this coincidence seriously and only considers models where TB mixing is automatically a good first approximation. Alternatively one can assume that the agreement of the data with TB mixing is accidental. Indeed there are many models that fit the data and yet TB mixing does not play any role in their architecture \cite{review}. The TB mixing matrix suggests that mixing angles are independent of mass ratios (while for quark mixings relations like $\lambda_C^2\sim m_d/m_s$ are typical). In fact in the basis where charged lepton masses are 
diagonal, the effective neutrino mass matrix in the TB case is given by $m_{\nu}=U_{TB}\rm{diag}(m_1,m_2,m_3)U_{TB}^T$:
\begin{equation}
m_{\nu}=  \left[\frac{m_3}{2}M_3+\frac{m_2}{3}M_2+\frac{m_1}{6}M_1\right]~~~~~. 
\label{1k1}
\end{equation}
where:
\be
M_3=\left(\matrix{
0&0&0\cr
0&1&-1\cr
0&-1&1}\right),~~~~~
M_2=\left(\matrix{
1&1&1\cr
1&1&1\cr
1&1&1}\right),~~~~~
M_1=\left(\matrix{
4&-2&-2\cr
-2&1&1\cr
-2&1&1}\right).
\label{4k1}
\ee
The eigenvalues of $m_{\nu}$ are $m_1$, $m_2$, $m_3$ with eigenvectors $(-2,1,1)/\sqrt{6}$, $(1,1,1)/\sqrt{3}$ and $(0,1,-1)/\sqrt{2}$, respectively. The expression in eq.(\ref{1k1}) can be reproduced in models with sequential dominance or with form dominance, discussed by S. King and collaborators \cite{ski}. 

As we see the most general neutrino mass matrix corresponding to TB mixing, in the basis of diagonal charged leptons, is of the form:
\begin{equation}
m=\left(\matrix{
x&y&y\cr
y&x+v&y-v\cr
y&y-v&x+v}\right),
\label{gl21}
\end{equation}
This is a symmetric, 2-3 symmetric matrix with $a_{11}+a_{12}=a_{22}+a_{23}$.

We now discuss models that naturally produce TB mixing in first approximation. Discrete non abelian groups naturally emerge as suiTable flavour symmetries \cite{RMP}. In fact the TB mixing matrix immediately suggests rotations by fixed, discrete angles. In a series of papers \cite{TBA4,AFextra,AFmodular,AFL,afh,altveram} it has been pointed out that a broken flavour symmetry based on the discrete
group $A_4$ appears to provide a simplest realisation of this specific mixing pattern in Leading Order (LO). Other
solutions based on alternative discrete or  continuous flavour groups have also been considered \cite{RMP}, but the $A_4$ models have a very economical and attractive structure, e.g. in terms of group representations and of field content. 

We recall that $A_4$, the group of even permutations of 4 objects, can be generated by the two elements
$S$ and $T$ obeying the relations (a "presentation" of the group):
\be
S^2=(ST)^3=T^3=1~~~.
\label{$A_4$}
\ee
The 12 elements of $A_4$  are obtained as:
$1$, $S$, $T$, $ST$, $TS$, $T^2$, $ST^2$, $STS$, $TST$, $T^2S$, $TST^2$, $T^2ST$.
The inequivalent irreducible representations of $A_4$ are 1, 1', 1" and 3. It is immediate to see that one-dimensional unitary representations are
given by:
\be
\begin{array}{lll}
1&S=1&T=1\\
1'&S=1&T=e^{\dd i 4 \pi/3}\equiv\omega^2\\
1''&S=1&T=e^{\dd i 2\pi/3}\equiv\omega \label{s$A_4$}
\end{array}
\ee
The three-dimensional unitary representation, in a basis
where the element $T$ is diagonal, is given by:
\be
T=\left(
\begin{array}{ccc}
1&0&0\\
0&\omega^2&0\\
0&0&\omega
\end{array}
\right),~~~~~~~~~~~~~~~~
S=\frac{1}{3}
\left(
\begin{array}{ccc}
-1&2&2\cr
2&-1&2\cr
2&2&-1
\end{array}
\right)~~~.
\label{ST}
\ee

Note that the generic mass matrix for TB mixing in eq.(\ref{gl21}) can be specified as the most general matrix that is invariant under $\mu-\tau$ symmetry, implemented by the unitary matrix  $A_{\mu \tau}$:
\be
A_{\mu \tau}=\left(
\begin{array}{ccc}
1&0&0\\
0&0&1\\
0&1&0
\end{array}
\right)
\label{Amutau}
\ee
and under the $S$ transformation:
\bea
m=SmS,~~~~~m=A_{\mu \tau}mA_{\mu \tau}~~\label{inv}
\eea
where S is given in eq.(\ref{ST}).
The $m$ mass matrix of the previous example is derived in the basis where charged leptons are diagonal. It is useful to consider the product $m^2=m_e^\dagger m_e$, where $m_e$ is the charged lepton mass matrix (defined as $\overline \psi_R m_e \psi_L$), because this product transforms as $m'^2=U_e^\dagger m^2 U_e$, with $U_e$ the unitary matrix that rotates the left-handed (LH) charged lepton fields. The most general diagonal $m^2$ is invariant under a diagonal phase matrix with 3 different phase factors:
\beq
m_e^\dagger m_e= T^\dagger m_e^\dagger m_e T
\label{Tdiag}
\eeq
and conversely a matrix $m_e^\dagger m_e$ satisfying the above requirement is diagonal. If $T^n=1$
the matrix $T$ generates a cyclic group $Z_n$.
The simplest case is $n=3$, which corresponds to $Z_3$ (but $n>3$ is equally possible) and to the $T$ matrix in eq.(\ref{ST}).

We can now see why $A_4$ works for TB mixing. It works because $S$ and $T$ are matrices of $A_4$ (in fact they satisfy eqs.(\ref{$A_4$})).  One could object that the matrix $A_{23}$ is not an element of $A_4$ 
(because the 2-3 exchange is an odd permutation). 
But it can be shown that in $A_4$ models the 2-3 symmetry is maintained by imposing that there are no flavons transforming as $1'$ or $1''$ that break $A_4$ with two different VEV's (in particular one can assume that there are no flavons in the model transforming as $1'$ or $1''$).

The group $A_4$ has two obvious subgroups: $G_S$, which is a reflection subgroup
generated by $S$ and $G_T$, which is the group generated by $T$, which is isomorphic to $Z_3$.
If the flavour symmetry associated to $A_4$ is broken by the VEV of a triplet
$\varphi=(\varphi_1,\varphi_2,\varphi_3)$ of scalar fields,
there are two interesting breaking pattern. The VEV
\be
\langle\varphi\rangle=(v_S,v_S,v_S)
\label{unotre}
\ee
breaks $A_4$ down to $G_S$, while
\be
\langle\varphi\rangle=(v_T,0,0)
\label{unozero}
\ee
breaks $A_4$ down to $G_T$. We have seen that $G_S$ and $G_T$ are the relevant low-energy symmetries
of the neutrino and the charged-lepton sectors, respectively. Indeed we have shown that the TB mass matrix is invariant under $G_S$
and, for charged leptons, a diagonal  $m_e^\dagger m_e$ is invariant under $G_T$. 
A crucial part of all serious A4 models is the dynamical generation of this alignment in a natural way.
In most of the models $A_4$ is accompanied by additional flavour symmetries, either discrete like $Z_N$ or continuous like U(1), which are necessary to eliminate unwanted couplings, to ensure the needed vacuum alignment and to reproduce the observed mass hierarchies. In the leading approximation $A_4$ models lead to exact TB mixing.  Given the set of flavour symmetries and having specified the field content, the non leading corrections to the TB mixing pattern arising from higher dimensional effective operators can be evaluated in a well defined expansion. In the absence of specific dynamical tricks, in a generic model, all the three mixing angles receive corrections of the same order of magnitude. Since the experimentally allowed departures of $\theta_{12}$ from the TB value $\sin^2{\theta_{12}}=1/3$ are small, at most of $\mathcal{O}(\lambda_C^2)$, with $\lambda_C$ the Cabibbo angle, it follows that, in these models, both $\theta_{13}$ and the deviation of $\theta_{23}$ from the maximal value are expected to also be at most of $\mathcal{O}(\lambda_C^2)$ (note that $\lambda_C$ is a convenient hierarchy parameter not only for quarks but also in the charged lepton sector with $m_\mu/m_\tau \sim0.06 \sim \lambda_C^2$ and $m_e/m_\mu \sim 0.005\sim\lambda_C^{3-4}$). A value of $\theta_{13} \sim \mathcal{O}(\lambda_C^2)$ is within the sensitivity of the experiments which are now in preparation and will take data in the near future. Explicit realizations of models for TB mixing based on $A_4$ can be found, for example, in \cite{AFextra,AFmodular,AFL,afh,altveram}. The possible origin of $A_4$ from a deeper level of the theory has been discussed in the context of extra dimensions and orbifolding \cite{AFL}, \cite{Adul} or as related to the fact that $A_4$ is a subgroup of the modular group \cite{AFmodular}, which plays a role in string theory.

While $A_4$ is the minimal flavour group leading to TB mixing, alternative flavour groups have been studied in the literature and can lead to interesting variants with some specific features \cite{RMP}. 

Recently, in ref. \cite{lam}, the claim was made that, in order to obtain the TB mixing "without fine tuning", the finite group must be $S_4$ or a larger group containing $S_4$. For us this claim is not well grounded being based on an abstract mathematical criterium for a natural model (see also \cite{gri}). For us a physical field theory model is natural if the interesting results are obtained from the most general lagrangian compatible with the stated symmetry and the specified representation content for the flavons. For example, we obtain from $A_4$ (which is a subgroup of $S_4$) a natural (in our sense) model for the TB mixing by simply not including symmetry breaking flavons transforming like the $1'$ and the $1''$ representations of $A_4$. This limitation on the transformation properties of the flavons is not allowed by the rules specified in ref. \cite{lam} which demand that the symmetry breaking is induced by all possible kinds of flavons (note that, according to this criterium, the SM of electroweak interactions would not be natural because only Higgs doublets are introduced!). Rather, for naturalness we also require that additional physical properties like the VEV alignment or the hierarchy of charged lepton masses also follow from the assumed symmetry and are not obtained by fine tuning parameters: for this actually $A_4$ can be more effective than $S_4$ because it possesses three different singlet representations 1, $1'$ and $1''$ which leads to unrelated masses for the three charged leptons.  Models of neutrino mixing based on $S_4$ have in fact been studied (see, for example, \cite{s4}).

\section{$A_4$, quarks and GUT's}

Much attention  has been devoted to the question whether models with TB mixing in the neutrino sector can be  suitably extended to also successfully describe the observed pattern of quark mixings and masses and whether this more complete framework can be made compatible with (supersymmetric) SU(5) or SO(10) Grand Unification. 

The simplest attempts of directly extending models based on $A_4$ to quarks have not been  satisfactory.
At first sight the most appealing
possibility is to adopt for quarks the same classification scheme under $A_4$ that one has
used for leptons (see, for example,  \cite{AFmodular}). Thus one tentatively assumes that LH quark doublets $Q$ transform
as a triplet $3$, while the  antiquarks $(u^c,d^c)$,
$(c^c,s^c)$ and $(t^c,b^c)$ transform as $1$, $1''$ and $1'$, respectively. This leads to $V_u=V_d$ and to the identity matrix for $V_{CKM}=V_u^\dagger V_d$ in the lowest approximation. This at first appears as very promising: a LO approximation where neutrino mixing is TB and $V_{CKM}=1$ is a very good starting point. But there are some problems. First, the corrections 
to $V_{CKM}=1$ turn out to be strongly constrained by the leptonic sector, because lepton mixing angles are very close to the TB values, and, in the simplest models, this constraint leads to a too small $V_{us}$
(i.e. the Cabibbo angle is rather large in comparison to the allowed shifts from the TB mixing angles). Also in these models, the quark classification which leads to $V_{CKM}=1$ is not compatible with $A_4$ commuting with SU(5). 
An additional consequence of the above assignment is that the top quark mass arises from a non-renormalizable dimension-5 operator. In that case, to reproduce the top mass, we need 
to compensate the cutoff suppression by some extra dynamical mechanism. Alternatively, we have to introduce a separate symmetry breaking parameter for the quark sector, sufficiently close to the cutoff
scale.

Due to this, larger discrete groups have been considered for the description of quarks.
A particularly appealing set of models is based on the discrete group $T'$, the double covering group of $A_4$ \cite{Tpr}, \cite{Feruglio:2007}, \cite{Chen:2007}. The 
representations of $T'$ are those of $A_4$ plus three independent doublets 2, $2'$ and $2''$. The doublets are interesting for the classification of the first two generations of quarks \cite{su2}. For example, in ref. \cite{Feruglio:2007} a viable description was obtained, i.e. in the leptonic sector the predictions of the $A_4$ model are maintained, while the $T'$ symmetry plays an essential role for reproducing the pattern of quark mixing. But, again, the classification adopted in this model is not compatible with Grand Unification.

As a result, the group $A_4$ was considered by many authors to be too
limited to also describe quarks and to lead to a grand unified
description. It has been recently shown \cite{afh} that this negative attitude
is not justified and that it is actually possible to construct a
viable model based on $A_4$ which leads to a grand
unified theory (GUT) of quarks and leptons with TB mixing
for leptons and with quark (and charged lepton) masses and mixings compatible with experiment. At the same time this model offers an example of an
extra dimensional SU(5) GUT in which a description of all fermion masses
and mixings is accomplished.  The
formulation of SU(5) in extra dimensions has the usual advantages of
avoiding large Higgs representations to break SU(5) and of solving the
doublet-triplet splitting problem.  The choice of the transformation properties of the two
Higgses $H_5$ and $H_{\overline{5}}$ has a special role in this model. They are chosen to transform 
as two different $A_4$ singlets
$1$ and $1'$. As a consequence, mass terms for the Higgs colour
triplets are  not directly allowed and their masses are
introduced by orbifolding, \`{a} la Kawamura \cite{Kawamura:2001}.  In this model, proton
decay is dominated by gauge vector boson exchange giving rise to
dimension-6 operators, while the usual contribution of dimension-5 operators is forbidden by the selection rules of the model. Given the large $M_{GUT}$ scale of SUSY models and the relatively huge theoretical uncertainties, the decay rate is within the present experimental limits.
A see-saw realization
in terms of an $A_4$ triplet of RH neutrinos $\nu^c$ ensures the
correct ratio of light neutrino masses with respect to the GUT
scale. In this model extra dimensional effects directly
contribute to determine the flavour pattern, in that the two lightest
tenplets $T_1$ and $T_2$ are in the bulk (with a doubling $T_i$ and
$T'_i$, $i=1,2$ to ensure the correct zero mode spectrum), whereas the
pentaplets $F$ and $T_3$ are on the brane. The hierarchy of quark and
charged lepton masses and of quark mixings is determined by a
combination of extra dimensional suppression factors and of $U(1)_{FN}$ charges, both of which only apply to the first two
generations, while the neutrino mixing angles
derive from $A_4$ in the usual way. If the extra dimensional suppression factors and the $U(1)_{FN}$ charges are switched off, only the third generation masses of quarks and charged leptons survive. Thus the charged fermion mass matrices are nearly empty in this limit (not much of $A_4$ effects remain) and the quark mixing angles are determined by the small corrections induced by the above effects. The model is natural, since most of the
small parameters in the observed pattern of masses and mixings as well
as the necessary vacuum alignment are  justified by the symmetries of
the model. However, in this case, like in all models based on $U(1)_{FN}$, the number of $\mathcal{O}(1)$ parameters is larger than the number of measurable quantities, so that in the quark sector the model can only account for the orders of magnitude (measured in terms of powers of an expansion parameter) and not for the exact values of mass ratios and mixing angles. A moderate fine tuning is only needed to enhance the Cabibbo mixing angle between the first two generations, which would generically be of $\mathcal{O}(\lambda_C^2)$. 

The problem of constructing GUT models based on  $SU(5)\otimes G_f$ or $SO(10)\otimes G_f$ with approximate TB mixing in the leptonic sector has  also been considered by many authors. Examples of models  based on $A_4$ are \cite{a4gut}.  An interesting model based on  $SU(5)\otimes T'$ is discussed in ref. \cite{Chen:2007}.  Recently some GUT models based on SU(5)$\times S_4$ have
appeared \cite{ishi}. 
As for the models based on $SO(10)\otimes G_f$  recent examples with $G_f=S_4$ are \cite{Dutta:2009} and $G_f=PSL_2(7)$ \cite{King:2009a}. Clearly the case of $SO(10)$ is even more difficult than that of $SU(5)$ because the neutrino sector is tightly related to that of quarks and charged leptons as all belong to the 16 of $SO(10)$ (for a discussion of $SO(10)\otimes A_4$ models, see \cite{Bazzocchi:2008b}). 
In our opinion most of the models are incomplete (for example, the crucial issue of VEV alignment is not really treated in depth as it should) and/or involve a number of unjustified steps and ad-hoc fine tuning of parameters. In particular, the problem of constructing a satisfactory natural model based on $SO(10)$ with built-in TB mixing at the LO approximation, remains open.

\section{Bimaximal Mixing and S4}

\begin{figure}
\centering
\includegraphics [width=10.0 cm]{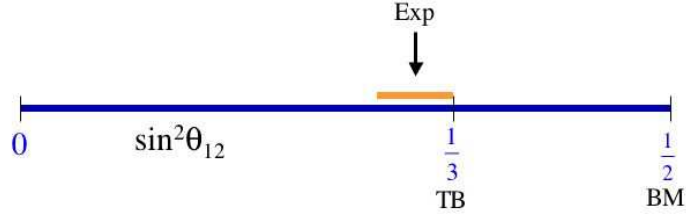}    
\caption{The values of $\sin^2{\theta_{12}}$ for TB o BM mixing are compared with the data}
\label{TBBM}
\end{figure}

Alternatively one can assume that the agreement of TB mixing with the data is accidental. Indeed there are many models that fit the data and yet TB mixing does not play a role in their architecture. However, in most cases, for this type of models different mixing angles could also be accommodated by simply varying the fitted values of the parameters. Assuming that the agreement of TB mixing with the data is accidental, we observe that the present data do not exclude a larger value for $\theta_{13}$, $\theta_{13} \sim \mathcal{O}(\lambda_C)$, than generally implied by models with approximate TB mixing (typically $\theta_{13} \sim \mathcal{O}(\lambda_C^2)$). In fact, two recent analyses of the available data lead to
$\sin^2{\theta_{13}}=0.016\pm0.010$ at 1$\sigma$ \cite{FogliIndication} and $\sin^2{\theta_{13}}=0.013^{+0.013}_{-0.009}$ at 1$\sigma$ \cite{MaltoniIndication}, which are compatible with both options. If experimentally it is found that $\theta_{13}$ is near its present upper bound, this could be interpreted as an indication that the agreement with the TB mixing is accidental. Then a scheme where instead the Bimaximal (BM) mixing is the correct first approximation could be relevant. The BM mixing matrix is given by:
\begin{equation}
U_{BM}= \left(\matrix{
\dd\frac{1}{\sqrt 2}&\dd-\frac{1}{\sqrt 2}&0\cr
\dd\frac{1}{2}&\dd\frac{1}{2}&-\dd\frac{1}{\sqrt 2}\cr
\dd\frac{1}{2}&\dd\frac{1}{2}&\dd\frac{1}{\sqrt 2}}\right)\;.
\label{21}
\end{equation}
 A comparison of the TB or BM mixing values with the data on $\sin^2{\theta_{12}}$ is shown in Fig. (\ref{TBBM}). 

In the BM scheme $\tan^2{\theta_{12}}= 1$, to be compared with the latest experimental
determination:  $\tan^2{\theta_{12}}= 0.45\pm 0.04$ (at $1\sigma$) \cite{FogliIndication,MaltoniIndication}, so that a rather large non leading correction is needed such that $\tan^2{\theta_{12}}$ is modified by terms of $\mathcal{O}(\lambda_C)$. This is in line with the well known empirical observation that $\theta_{12}+\lambda_C\sim \pi/4$, a relation known as quark-lepton complementarity \cite{compl}, or similarly $\theta_{12}+\sqrt{m_\mu/m_\tau} \sim \pi/4$. No compelling model leading, without parameter fixing, to the exact complementarity relation has been produced so far. Probably the exact complementarity relation is to be replaced with something like $\theta_{12}+\mathcal{O}(\lambda_C)\sim \pi/4$ or $\theta_{12}+\mathcal{O}(m_\mu/m_\tau)\sim \pi/4$ (which we could call "weak" complementarity), as in models where the large $\nu$ mixings arise from the diagonalisation of charged leptons.
Along this line of thought, the expertise acquired with non Abelian finite flavour groups can be used to construct a model \cite{S4us} based on the permutation group $S_4$ which naturally leads to the BM mixing at LO. The model is supersymmetric in 4 space-time dimensions and the complete flavour group is $S_4\times Z_4 \times U(1)_{FN}$. In LO, the charged leptons are diagonal and hierarchical and the light neutrino mass matrix, after see-saw, leads to the exact BM mixing. The model is built in such a way that the dominant corrections to the BM mixing pattern, arising from higher dimensional operators in the superpotential,  only arise from the charged lepton sector and naturally inherit $\lambda_C$ as the relevant expansion parameter. As a result the mixing angles deviate from the BM values by terms of  $\mathcal{O}(\lambda_C)$ (at most), and weak complementarity holds. A crucial feature of the model is that only $\theta_{12}$ and $\theta_{13}$ are corrected by terms of $\mathcal{O}(\lambda_C)$ while $\theta_{23}$ is unchanged at this order (which is essential for the model to agree with the present data). 

\section{Conclusion}

In the last decade we have learnt a lot about neutrino masses and mixings.  A list of important conclusions have been reached. Neutrinos are not all massless but their masses are very small. Probably masses are small because neutrinos are Majorana particles
with masses inversely proportional to the large scale M of lepton number violation. It is quite remarkable that M is empirically not far from $M_{GUT}$, so that
neutrino masses fit well in the SUSY GUT picture. Also out of equilibrium decays with CP and L violation of heavy RH neutrinos can produce a B-L asymmetry, then converted near the weak scale by instantons into an amount of B asymmetry compatible with observations (baryogenesis via leptogenesis) \cite{buch}.  It has been established that neutrinos are not a significant component of dark matter in the Universe. We have also understood there there is no contradiction between large neutrino mixings and small quark mixings, even in the context of GUTs.  

This is a very impressive list of achievements. Coming to a detailed analysis of neutrino masses and mixings a very long collection of models have been formulated over the years. 
With continuous improvement of the data and more precise values of the mixing angles most of the models have been discarded by experiment. By now, besides the detailed knowledge of the entries of the $V_{CKM}$ matrix we also have a reasonable determination of the neutrino mixing matrix $U_{P-MNS}$. It is a fact that, to a precision comparable with the measurement accuracy, the TB mixing pattern is well approximated by the data  (see Fig. (2)). If this experimental result is not a mere accident but a real indication that a dynamical mechanism is at work to guarantee the validity of TB mixing in the leading approximation, corrected by small non leading terms, then non abelian discrete flavour groups emerge as the main road to an understanding of this mixing pattern. Indeed the entries of the TB mixing matrix are clearly suggestive of "rotations" by simple, very specific angles. It is remarkable that neutrino and  quark mixings have such a different qualitative pattern. In the near future the improved experimental precision on neutrino mixing angles, in particular on $\theta_{13}$, could make the case for TB mixing stronger and then, as a consequence, also the case for discrete flavour groups would be strenghtened. An obvious question is whether some additional indication for discrete flavour groups can be obtained by considering the extension of the models to the quark sector, perhaps in a Grand Unified context. The answer appears to be that, while the quark masses and mixings can indeed be reproduced in models where TB mixing is realized in the leptonic sector through the action of discrete groups, there are no specific additional hints in favour of discrete groups that come from the quark sector. Further important input could come from the LHC. In fact, new physics at the weak scale could have important feedback on the physics of neutrino masses and mixing.

In conlusion, one could have imagined that neutrinos would bring a decisive boost towards the formulation of a comprehensive understanding of fermion masses and mixings. In reality it is frustrating that no real illumination was sparked on the problem of flavour. We can reproduce in many different ways the observations but we have not yet been able to single out a unique and convincing baseline for the understanding of fermion masses and mixings. In spite of many interesting ideas and the formulation of many elegant models the mysteries of the flavour structure of the three generations of fermions have not been much unveiled.


\begin{thebibliography}{99}

\bibitem{review}
G.~Altarelli and F.~Feruglio,
  New J.\ Phys.\  {\bf 6} (2004) 106
  [ArXiv:hep-ph/0405048];
G.~Altarelli, [ArXiv:0905.3265].

\bibitem{rev2}  
R. N. Mohapatra and A. Y. Smirnov,
Ann. Rev. Nucl. Part. Sci. 56, 569 (2006)
[ArXiv:hep-ph/0603118];
W. Grimus, PoS P2GC:001,2006. 
[ArXiv:hep-ph/0612311];
M.~C.~Gonzalez-Garcia and M.~Maltoni,
  Phys.\ Rept.\  {\bf 460} (2008) 1
  [ArXiv:0704.1800 ].
  
\bibitem{Mini} The MiniBooNE Collaboration
Phys.Rev.Lett. {\bf105} 181801 (2010), [ArXiv:1007.1150].

\bibitem{FogliIndication}
G.~L.~Fogli, E.~Lisi, A.~Marrone, A.~Palazzo and A.~M.~Rotunno,
  Phys.\ Rev.\ Lett.\  {\bf 101} (2008) 141801,
  [ArXiv:0806.2649 ].
G.~L.~Fogli, E.~Lisi, A.~Marrone, A.~Palazzo and A.~M.~Rotunno,
  [ArXiv:0809.2936].

\bibitem{MaltoniIndication}
T.~Schwetz, M.~Tortola and J.~W.~F.~Valle,
  New J.\ Phys.\  {\bf 10} (2008) 113011
  [ArXiv:0808.2016 ];
M.~Maltoni and T.~Schwetz,
  [ArXiv:0812.3161].
  
\bibitem{fo} G.~L.~Fogli et al, [ArXiv:0805.2517].

\bibitem{seesaw}
P. Minkowski, Phys. Letters B67 (1977)421; T. Yanagida, in \emph{Proc.\ of the Workshop
on Unified Theory and Baryon Number in the Universe}, KEK, March 1979;
S. L. Glashow, in ``Quarks and Leptons'', Carg\`ese, ed. M. L\'evy et al.,
Plenum, 1980 New York, p. 707;
M. Gell-Mann, P. Ramond and R. Slansky,
in \emph{Supergravity}, Stony Brook, Sept 1979;
R.~N.~Mohapatra and G.~Senjanovic,
Phys.\ Rev.\ Lett.\  {\bf 44} (1980) 912.

\bibitem{fsv} F. Feruglio, A. Strumia and F Vissani,
Nucl.Phys.B637:345-377,2002., Addendum-ibid.B659:359-362, 2003.
[ArXiv:hep-ph/0201291].

\bibitem{tro} For a review see, for example,: M.~Trodden,
Rev.\ Mod.\ Phys.\  {\bf 71}, 1463 (1999).

\bibitem{buch} For a review see for example: W. Buchmuller, R.D. Peccei and T. Yanagida, Ann.Rev.Nucl.Part.Sci.55:311-355,2005.
 [ArXiv:hep-ph/0502169].
 
\bibitem{hps}
P.~F.~Harrison, D.~H.~Perkins and W.~G.~Scott,
Phys.\ Lett.\ B {\bf 530} (2002) 167,
[ArXiv:hep-ph/0202074];
P.~F.~Harrison and W.~G.~Scott,
Phys.\ Lett.\ B {\bf 535} (2002) 163,
[ArXiv:hep-ph/0203209]; Phys.\ Lett.\ B {\bf 547} (2002) 219,
[ArXiv:hep-ph/0210197]; Phys.\ Lett.\ B {\bf 557} (2003) 76,
[ArXiv:hep-ph/0302025];
ArXiv:hep-ph/0402006;
ArXiv:hep-ph/0403278.

\bibitem{ski} 
S.~F.~King, ArXiv:0904.3255 and references therein.

\bibitem{RMP} For a recent review and a list of references, see  
G.~Altarelli and F.~Feruglio, Rev.\ Mod.\ Phys. in press, ArXiv:1002.0211.

\bibitem{TBA4}
E.~Ma and G.~Rajasekaran,
  Phys.\ Rev.\ D {\bf 64} (2001) 113012
  [ArXiv:hep-ph/0106291];
E.~Ma,
  Mod.\ Phys.\ Lett.\ A {\bf 17} (2002) 627
  [ArXiv:hep-ph/0203238].
K.~S.~Babu, E.~Ma and J.~W.~F.~Valle,
  Phys.\ Lett.\ B {\bf 552} (2003) 207
  [ArXiv:hep-ph/0206292];
M.~Hirsch, J.~C.~Romao, S.~Skadhauge, J.~W.~F.~Valle and A.~Villanova del Moral,
  ArXiv:hep-ph/0312244;
  Phys.\ Rev.\  D {\bf 69} (2004) 093006
  [ArXiv:hep-ph/0312265];
E.~Ma,
  Phys.\ Rev.\ D {\bf 70} (2004) 031901;
  Phys.\ Rev.\ D {\bf 70} (2004) 031901
  [ArXiv:hep-ph/0404199];
  New J.\ Phys.\  {\bf 6} (2004) 104
  [ArXiv:hep-ph/0405152];
  ArXiv:hep-ph/0409075;
S.~L.~Chen, M.~Frigerio and E.~Ma,
  Nucl.\ Phys.\  B {\bf 724} (2005) 423
  [ArXiv:hep-ph/0504181];
E.~Ma,
  Phys.\ Rev.\  D {\bf 72} (2005) 037301
  [ArXiv:hep-ph/0505209];
M.~Hirsch, A.~Villanova del Moral, J.~W.~F.~Valle and E.~Ma,
   Phys.\ Rev.\  D {\bf 72} (2005) 091301
   [Erratum-ibid.\  D {\bf 72} (2005) 119904]
   [ArXiv:hep-ph/0507148].
K.~S.~Babu and X.~G.~He,
  ArXiv:hep-ph/0507217;
E.~Ma,
  Mod.\ Phys.\ Lett.\ A {\bf 20} (2005) 2601
  [ArXiv:hep-ph/0508099];
A.~Zee,
  Phys.\ Lett.\ B {\bf 630} (2005) 58
  [ArXiv:hep-ph/0508278];
E.~Ma,
  Phys.\ Rev.\  D {\bf 73} (2006) 057304
  [ArXiv:hep-ph/0511133];
X.~G.~He, Y.~Y.~Keum and R.~R.~Volkas,
  JHEP {\bf 0604} (2006) 039
  [ArXiv:hep-ph/0601001];
B.~Adhikary, B.~Brahmachari, A.~Ghosal, E.~Ma and M.~K.~Parida,
  Phys.\ Lett.\ B {\bf 638} (2006) 345
  [ArXiv:hep-ph/0603059];
E.~Ma,
  Mod.\ Phys.\ Lett.\  A {\bf 21} (2006) 2931
  [ArXiv:hep-ph/0607190];
  Mod.\ Phys.\ Lett.\  A {\bf 22} (2007) 101
  [ArXiv:hep-ph/0610342];
L.~Lavoura and H.~Kuhbock,
  Mod.\ Phys.\ Lett.\  A {\bf 22} (2007) 181
  [ArXiv:hep-ph/0610050];
S.~F.~King and M.~Malinsky,
  Phys.\ Lett.\  B {\bf 645} (2007) 351
  [ArXiv:hep-ph/0610250];
M.~Hirsch, A.~S.~Joshipura, S.~Kaneko and J.~W.~F.~Valle,
   Phys.\ Rev.\ Lett.\  {\bf 99}, 151802 (2007)
   [ArXiv:hep-ph/0703046].
F.~Yin,
  Phys.\ Rev.\  D {\bf 75} (2007) 073010
  [ArXiv:0704.3827 ];
F.~Bazzocchi, S.~Kaneko and S.~Morisi,
  JHEP {\bf 0803} (2008) 063
  [ArXiv:0707.3032 ].
F.~Bazzocchi, S.~Morisi and M.~Picariello,
  Phys.\ Lett.\  B {\bf 659} (2008) 628
  [ArXiv:0710.2928 ];
M.~Honda and M.~Tanimoto,
  Prog.\ Theor.\ Phys.\  {\bf 119} (2008) 583
  [ArXiv:0801.0181 ];
B.~Brahmachari, S.~Choubey and M.~Mitra,
  Phys.\ Rev.\  D {\bf 77} (2008) 073008
  [Erratum-ibid.\  D {\bf 77} (2008) 119901]
  [ArXiv:0801.3554 ];
B.~Adhikary and A.~Ghosal,
  Phys.\ Rev.\  D {\bf 78} (2008) 073007
  [ArXiv:0803.3582 ];
M.~Hirsch, S.~Morisi and J.~W.~F.~Valle,
  Phys.\ Rev.\  D {\bf 78} (2008) 093007
  [ArXiv:0804.1521 ].
P.~H.~Frampton and S.~Matsuzaki,
  ArXiv:0806.4592 ;
  C. ~Csaki, C.~ Delaunay, C. ~Grojean, Y.~Grossman
   ArXiv:0806.0356 ;
F.~Feruglio, C.~Hagedorn, Y.~Lin and L.~Merlo,
  ArXiv:0807.3160 ;
F.~Bazzocchi, M.~Frigerio and S.~Morisi,
  ArXiv:0809.3573 ;
W.~Grimus and L.~Lavoura,
  ArXiv:0811.4766 ;
S.~Morisi,
  ArXiv:0901.1080 ;
 M.~C.~Chen and S.~F.~King,
  ArXiv:0903.0125 .
Y.~Lin,
Nucl.\ Phys.\  B {\bf 813}, 91 (2009)
[ArXiv:0804.2867 ];
ArXiv:0903.0831;
F.~del Aguila, A. Carmona  and J.~Santiago, ArXiv:1001.515
A.~Kadosh and E.~Pallante, ArXiv:1004.0321;
D.~Meloni, S.~Morisi and E.~Peinado, ArXiv:1011.1371;
E.~Peinado, ArXiv:1010.2614;
F.~Feruglio and A.~Paris, Nucl.Phys. {\bf B840} (2010) 405,
[ArXiv:1005.5526];

\bibitem{AFextra}
G.~Altarelli and F.~Feruglio,
  Nucl.\ Phys.\ B {\bf 720} (2005) 64
  [ArXiv:hep-ph/0504165].

\bibitem{AFmodular}
G.~Altarelli and F.~Feruglio,
  Nucl.\ Phys.\  B {\bf 741} (2006) 215
  [ArXiv:hep-ph/0512103].

\bibitem{AFL}
G.~Altarelli, F.~Feruglio and Y.~Lin,
  Nucl.\ Phys.\  B {\bf 775} (2007) 31
  [ArXiv:hep-ph/0610165].
  
\bibitem{afh}G.~Altarelli, F.~Feruglio and C.~Hagedorn,
  JHEP {\bf 0803} (2008) 052
  [ArXiv:0802.0090 ].

\bibitem{altveram} 
G.~Altarelli and D.~Meloni, ArXiv:0905.0620.

\bibitem{s4}
 R.N.~ Mohapatra, M.K.~ Parida and G.~Rajasekaran, Phys. Rev.  {\bf D69}
  (2004) 053007, [ArXiv:hep-ph/0301234];
C.~Hagedorn,  M.~Lindner and R.N.~ Mohapatra, JHEP  {\bf 06} (2006) 042,
[ArXiv:hep-ph/0602244];
Y.~Cai  and H.B~Yu,  Phys. Rev. {\bf D74} (2006) 115005,  [ArXiv:hep-ph/0608022];
E.~Ma, Phys. Lett. {\bf B632} (2006) 352; [ArXiv:hep-ph/0508231];
F.~Bazzocchi  and S.~Morisi, Phys. Rev.  {\bf D80} (2009) 096005;
 [ArXiv:0811.0345];
H.~Ishimori, Y.~Shimizu  and  M.~Tanimoto, Prog. Theor. Phys. {\bf 121}
 (2009) 769,   [ArXiv:0812.5031];
F.~Bazzocchi, L.~Merlo  and S.~Morisi, Nucl. Phys {\bf B816} ( 2009 ) 204,   [ArXiv:0901.2086];
Phys. Rev. {\bf D80} (2009) 053003,
 [ArXiv:0902.2849];
D.~Meloni, J. Phys. {\bf G37} (2010) 055201,[ArXiv: 0911.3591].
G.J.~Ding, Nucl. Phys.  {\bf B827} (2010) 82, [ArXiv:0909.2210];
 S.~Morisi and  E. Peinado, Phys. Rev. {\bf D81} (2010) 085015,
 [ArXiv:1001.2265];
 C.~Hagedorn, S.F.~ King and  C.~Luhn, ArXiv:1003.4249;
H.~Ishimori \textit{et al}, ArXiv:1004.5004.

\bibitem{Adul} 
A.~Adulpravitchai, A.~Blum and  M.~Lindner, JHEP {\bf 07} (2009) 053
ArXiv:0906.0468; 
A.~Adulpravitchai, M.A.~Schmidt, ArXiv:1001.3172;
T.J.~Burrows, S.F.~King, Nucl.Phys. {\bf B842} (2011) 107, 
[ArXiv:1007.2310].

\bibitem{lam}
  C.~S.~Lam, Phys.Rev.Lett.101:121602,2008, [ArXiv:0804.2622];
  Phys.\ Rev.\   {\bf D\,78}  (2008) 073015,
  [ArXiv:0809.1185].

\bibitem{gri} 
W.~Grimus, L.~Lavoura and P.~O.~ Ludl, 
J.Phys. {\bf G36} (2009) 15007, [ArXiv:0906.2689].

\bibitem{Tpr}
P.H.~Frampton and T.W.~ Kephart,   Int. J. Mod. Phys.  {\bf A10}
(1995) 4689, [ArXiv:hep-ph/9409330]; JHEP  {\bf 09} (2007) 110, [ArXiv:0706.1186];
A.~Aranda, C.D.~Carone  and  R.F.~ Lebed, Phys. Lett. {\bf B474} (2000) 170, 
[ArXiv:hep-ph/9910392]; Phys. Rev. {\bf D62} (2000) 016009, [ArXiv:hep-ph/0002044];
P.D.~Carr and P.H.~Frampton, ArXiv:hep-ph/0701034;
A.~Aranda, Phys. Rev. {\bf D76} (2007) 111301, [ArXiv:0707.3661];
P.H.~Frampton and S.~Matsuzaki, Phys. Lett. {\bf B679} (2009) 347, [ArXiv:0902.1140];
G.J.~ Ding, Phys. Rev. {\bf D78} (2008) 036011, [ArXiv:0803.2278];

\bibitem{Feruglio:2007} F.~Feruglio \textit{et al}, Nucl. Phys.  {\bf B775} (2007) 120, [ArXiv:hep-ph/0702194].

\bibitem{Chen:2007}
 M.C.~Chen and K.T.~Mahanthappa, Phys. Lett. {\bf B652} (2007) 34, ArXiv:0705.0714.

\bibitem{su2}
A.~Pomarol and D.~Tommasini,  Nucl. Phys. {\bf B466} ( 1996) 3,  [ArXiv:hep-ph/9507462];
R.~Barbieri,  G.R.~Dvali and L.~Hall, Phys. Lett. {\bf B377} (1996) 76, [ArXiv:hep-ph/9512388]
R.~Barbieri \textit{et al}, Nucl. Phys. {\bf B493} (1997) 3, [ArXiv:hep-ph/9610449];
R.~Barbieri, L.~Hall and A.~Romanino, Phys. Lett. {\bf B401} (1997) 47, [ArXiv:hep-ph/9702315].

\bibitem{Kawamura:2001}
Y. Kawamura, Prog. Theor. Phys.  {\bf 105} (2001) 999, [ArXiv:hep-ph/0012125].

\bibitem{a4gut}
E.~Ma,
  Mod.\ Phys.\ Lett.\  A {\bf 20} (2005) 2767, [ArXiv:hep-ph/0506036];
E.~Ma, H.~Sawanaka and M.~Tanimoto,
  Phys.\ Lett.\  B {\bf 641} (2006) 301
  [ArXiv:hep-ph/0606103];
E.~Ma,
  Mod.\ Phys.\ Lett.\  A {\bf 21} (2006) 2931
  [ArXiv:hep-ph/0607190];
S.~Morisi, M.~Picariello and E.~Torrente-Lujan,
  Phys.\ Rev.\  D {\bf 75} (2007) 075015
  [ArXiv:hep-ph/0702034];
  W.~Grimus and H.~Kuhbock,
  ArXiv:0710.1585; 
P.~Ciafaloni \textit{et al}, Phys.
  Rev. {\bf D79} (2009) 116010, [ArXiv:0901.2236];
F.~Bazzocchi \textit{et al}, J. Phys. 
  {\bf G36} (2009) 015002, [ArXiv:0802.1693];
S.~Antusch,  S.F.~King and M.~Spinrath, ArXiv:1005.0708].

\bibitem{ishi}
C.~Hagedorn, S.F.~King and C. Luhn, ArXiv:1003.4249];
H.~Ishimori\textit{et al}, ArXiv:1004.5004].

\bibitem{Dutta:2009}
 B.~Dutta, Y.~Mimura and R.N.~ Mohapatra, Phys. Rev. {\bf D80} (2009) 095021,
 [ArXiv:0910.1043]; JHEP {\bf 05} (2010) 034, [ArXiv:0911.2242];

\bibitem{King:2009a}
S. F.~King  and C. Luhn, [ArXiv:0912.1344].

\bibitem{Bazzocchi:2008b}
F.~Bazzocchi, F.~Frigerio and S.~Morisi, Phys. Rev. {\bf D78} (2008)116018, 
[ArXiv:0809.3573].

\bibitem{compl}
M.~Raidal,
  Phys.\ Rev.\ Lett.\  {\bf 93} (2004) 161801
  [ArXiv:hep-ph/0404046];
H.~Minakata and A.~Y.~Smirnov,
  Phys.\ Rev.\  D {\bf 70} (2004) 073009
  [ArXiv:hep-ph/0405088];
H.~Minakata,
ArXiv:hep-ph/0505262;
  P.~H.~Frampton and R.~N.~Mohapatra,
  JHEP {\bf 0501}, 025 (2005), hep-ph/0407139;
  J.~Ferrandis and S.~Pakvasa,
  Phys.\ Rev.\ D {\bf 71} (2005) 033004,
  hep-ph/0412038;
 S.~K.~Kang, C.~S.~Kim and J.~Lee,
  ArXiv:hep-ph/0501029;
  G.~Altarelli, F.~Feruglio and I.~Masina,
  Nucl.\ Phys.\  B {\bf 689} (2004) 157
  [ArXiv:hep-ph/0402155];
  N.~Li and B.~Q.~Ma,
  hep-ph/0501226;
  K.~Cheung, S.~K.~Kang, C.~S.~Kim and J.~Lee,
  hep-ph/0503122;
  Z.~z.~Xing,
  hep-ph/0503200;
  A.~Datta, L.~Everett and P.~Ramond,
  ArXiv:
  hep-ph/0503222;
T.~Ohlsson,
ArXiv:hep-ph/0506094;
S.~Antusch, S.~F.~King and R.~N.~Mohapatra,
ArXiv:hep-ph/0504007.;
M.~Lindner, M.~A.~Schmidt and A.~Y.~Smirnov,
ArXiv:hep-ph/0505067;
S.~F.~King,
JHEP {\bf 0508} (2005) 105
[ArXiv:hep-ph/0506297];
A.~Dighe, S.~ Goswami, and P.~Roy, Phys.Rev.{\bf D73} (2006) 07130,
[ArXiv:hep-ph/0602062];
B.~ C.~ Chauhan, M.~ Picariello, J.~ Pulido and E.~ Torrente-Lujan, Eur.Phys. {\bf J.C50}(2007) 573,
[ArXiv:hep-ph/0605032]
M.~A. ~Schmidt and A.~ Yu.~ Smirnov, Phys.Rev. {\bf D74}(2006)113003,
[ArXiv:hep-ph/0607232];
K.~ A~ Hochmuth and W.~ Rodejohann, Phys.Rev. {\bf D75}(2007) 073001,
[ArXiv:hep-ph/0607103];
F.~ Plentinger, G.~ Seidl and W.~ Winter, Nucl.Phys. {\bf B791} (2008) 60,
[ArXiv: hep-ph/0612169]; Phys.Rev. {\bf D76} (2007)113003,
[ArXiv:0707.2379 ].

\bibitem{S4us} 
G.~Altarelli, F.~Feruglio and L.~Merlo,
[ ArXiv:0903.1940]


\end{thebibliography}
\end{document}